\documentclass[aip, jmp, amsmath,twocolumn,amssymb,reprint,]{revtex4} 

\usepackage{graphicx,epstopdf}
\usepackage[usenames]{color}
\usepackage{latexsym}
\usepackage{hyperref}
\usepackage{bm}

\usepackage{docs}%
\usepackage{dcolumn}

\begin{document}
\title{Electrically detected magnetic resonance of neutral donors interacting with a two-dimensional electron gas}

\author{C. C. Lo$^1$}
\email[Both authors contributed to this work equally. Please contact the corresponding author under ]{cclo@eecs.berkeley.edu.}
\author{V. Lang$^2$}
\email[Both authors contributed to this work equally. Please contact the corresponding author under ]{cclo@eecs.berkeley.edu.}
\author{R. E. George$^3$}
\author{J. J. L. Morton$^{2,3}$}
\author{A. M. Tyryshkin$^4$}
\author{S. A. Lyon$^4$}
\author{J. Bokor$^{1}$}
\author{T. Schenkel$^{5}$}
\affiliation{$^1$Department of Electrical Engineering and Computer Sciences, University of California, Berkeley, California 94720, USA}
\affiliation{$^2$Department of Materials, University of Oxford, Oxford OX1 3PH, United Kingdom}
\affiliation{$^3$CAESR, Clarendon Laboratory, Department of Physics, University of Oxford, Oxford OX1 3PU, United Kingdom}
\affiliation{$^4$Department of Electrical Engineering, Princeton University, New Jersey 08544, USA}
\affiliation{$^5$Accelerator and Fusion Research Division, Lawrence Berkeley National Laboratory, Berkeley, California 94720, USA}
\date{\today}
\begin{abstract}
We have measured the electrically detected magnetic resonance of channel-implanted donors in silicon field-effect transistors in resonant X- ($9.7\:$GHz) and W-band ($94\:$GHz) microwave cavities, with corresponding Zeeman fields of $0.35\:$T and $3.36\:$T, respectively. It is found that the conduction electron resonance signal increases by two orders of magnitude from X- to W-band, while the hyperfine-split donor resonance signals are enhanced by over one order of magnitude. We rule out a bolometric origin of the resonance signals, and find that direct spin-dependent scattering between the two-dimensional electron gas and neutral donors is inconsistent with the experimental observations. We propose a new polarization transfer model from the donor to the conduction electrons as the main contributer to the spin resonance signals observed.
\end{abstract}



\maketitle

Electrical spin-state detection for solid-state qubits requires a detection channel formed by conduction electrons in close proximity to the qubit. For electron spin qubits, the detection channels usually consist of quantum point contacts or single electron transistors, which are sensitive to the electrostatic environment nearby and able to detect the spin-dependent occupancies of electrons at the qubit site \cite{virjen00,elzerman04, xiao10, morello10}. Alternatively, for nuclear spin qubits such as shallow donors in silicon \cite{kane98}, it was proposed that conduction electrons interacting {\it directly} with the neutral donors can be used for nuclear spin-state readout \cite{sarovar08}, as the conduction and neutral donor electrons undergo spin-dependent scattering \cite{ghosh92, lo07, vanbeveren08, huebl09, desousa09}. 
\begin{figure}[h]
\includegraphics[width=3.5in]{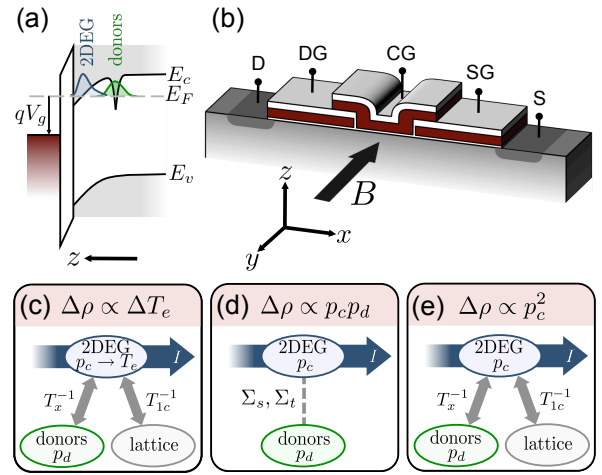}
\caption{\label{fig1}{(a) Energy-band diagram of the MOS system. The overlap of the electron wavefunctions between the 2DEG and neutral donor is also illustrated. (b) Schematic of the aFET used in this study, where the drain (D) and source (S) are separated by three gates (DG, CG and SG) forming the triple-gate structure. Phosphorus donors are present under all three gates while arsenic donors are only present under the center gate (CG) region. (c$-$e) Three possible EDMR mechanisms affecting the 2DEG detection channel current $I$ (blue arrow), and the expected change in 2DEG resistivity $\Delta\rho$ associated with each mechanism: 
(c) 2DEG bolometric heating, (d) spin-dependent scattering and (e) 2DEG polarization transfer. The grey arrows represent energy transfer between the systems, while the dashed line in (d) represents elastic scattering. See text for the definition of symbols and details of each mechanism.}}
\end{figure}
Donor-doped metal-oxide-semiconductor (MOS) devices provide an ideal platform for the detection of such an interaction, as the electronic wavefunction of neutral donors embedded in the device channel can overlap with the nearby gate-induced two-dimensional electron gas (2DEG) (Fig.\:\ref{fig1}(a)). 
The donor-2DEG interaction can be probed by electrically detected magnetic resonance (EDMR) experiments with the MOS system, as was first reported by Ghosh and Silsbee \cite{ghosh92}. However, the use of bulk-doped silicon with a relatively high donor concentration resulted in significant overlap between the donor and 2DEG electron resonance signals, complicating the analysis of the results. In addition, their measurements were limited to a low magnetic field of $\sim\:0.35\:$T. In this Letter, we clarify the mechanisms behind the EDMR signals of such donor-doped MOS devices by studying the change in EDMR signal intensities at different magnetic fields. We perform EDMR with accumulation-mode n-type field-effect transistors (aFETs) at Zeeman fields of approximately $3.36\:$T and compare it to low-field EDMR at $0.35\:$T. The low donor concentrations used in this work enable clear identifications of the 2DEG and donor contributions to the resonance signal. We will discuss our results in terms of (i) bolometric heating, (ii) spin-dependent scattering, and (iii) a polarization transfer from the donor to the 2DEG spin system. 

Bolometric heating of the 2DEG (Fig.\:\ref{fig1}(c)) can occur when the 2DEG kinetic energy, characterized by the orbital electron temperature $T_e$, rises as a result of an increase of the 2DEG electron spin temperature (i.e. a decrease in the 2DEG spin density polarization $p_c$) via spin-orbit interaction \cite{morigaki74}. The energy transfer from the 2DEG spins to the lattice occurs through $T_{1c}$ relaxation process and from donor spins through flip-flop $T_{1x}$ process via exchange scattering with 2DEG electrons. This bolometric response is expected to be enhanced at higher magnetic fields as the absorbed Zeeman energy on resonance is increased. 

Spin-dependent scattering (Fig.\:\ref{fig1}(d)) arises from a difference in the scattering cross sections $\Sigma_s$ and $\Sigma_t$ when the 2DEG and donor electrons form singlet ($s$) and triplet ($t$) pairs, respectively. 
In thermal equilibrium there is an excess of triplet pairs, and the number of singlet pairs is increased when either the donor or 2DEG electron spins are resonantly excited. This leads to a change in the 2DEG mobility with an expected fractional change in sample resistivity of $\Delta\rho/\rho_0 \propto p_c p_d$ under full power saturation, where $\rho_0$ is the 2DEG resistivity in thermal equilibrium, $p_c$ and $p_d$ the spin density polarizations for the conduction and donor electrons, respectively \cite{ghosh92}. For an ideal 2DEG, $p_c \propto g\mu_B B$, where $g$ is the Land$\mathrm{\acute{e}}$ $g$-factor, $\mu_B$ the Bohr magneton and $B$ the magnetic field. For donors, $p_d = $tanh$(g\mu_B B/k_BT)$,  with $k_B$ the Boltzmann constant and $T$ the temperature. This implies that the 2DEG and donor resonance signals should have the same magnetic field dependence, as only the product of the polarizations are measured under the spin-dependent scattering mechanism. 

The third mechanism we consider here results from the polarization dependence of the 2DEG resistivity \cite{abrahams01,pudalov02,okamoto04}, as was found to be the case for the EDMR of high mobility silicon 2DEGs \cite{graeff99, matsunami06}. Under this mechanism, donor electrons can contribute to a resonant change in 2DEG resistivity as the donor polarization is transferred to the 2DEG spin system via exchange scattering (Fig.\:\ref{fig1}(e)). The observation of this effect is possible only if the spin-orbit coupling is weak and $T_e$ is not perturbed excessively, as the bolometric response will dominate otherwise. 
The three mechanisms discussed above will form the basis for the detailed discussion of our results later on.

A schematic of the aFET used in this study is shown in Fig.\:\ref{fig1}(b). The device was fabricated on 1$\:\mu$m thick 99.99$\%$ isotopically purified 28-silicon ($^{28}$Si), grown epitaxially on a high resistivity natural silicon substrate. The aFET has a triple-gate geometry where the channel can be separated into three regions: two $60\:\mu$m long side gates and one $40\:\mu$m long center gate, with a channel width of $40\:\mu$m and $20\:$nm gate oxide thickness throughout. The top $1\:\mu$m layer of the silicon substrate is doped with $3\times10^{16}\:$cm$^{-3}$ phosphorus ($^{31}$P) donors, while the center region received an additional implantation of arsenic ($^{75}$As) donors at $50\:$keV and a dose of $4\times10^{11}\:$cm$^{-2}$. For this study all three gates are biased together and the whole device is considered as a simple three-terminal MOSFET. Secondary ion mass spectroscopy (SIMS) was carried out to determine the post-processing donor concentrations under the channel. 
Both donor species exhibited pile-up behavior at the oxide interface, where the phosphorus and arsenic had peak concentrations of $1\times10^{17}\:$cm$^{-3}$ and $5.5\times10^{16}\:$cm$^{-3}$, respectively.
From the geometry of the device approximately $6\times10^{5}$ arsenic and $4\times10^{6}$ phosphorus donors reside within $10\:$nm of the oxide interface of the device channel, where they can interact with the 2DEG electrons directly. 
A silicon dioxide-aluminum stack, acting as a microwave shunt, is deposited over the sample in order to minimize microwave-induced rectification noise \cite{lo10}. 

\begin{figure}
\includegraphics[width=3.3in]{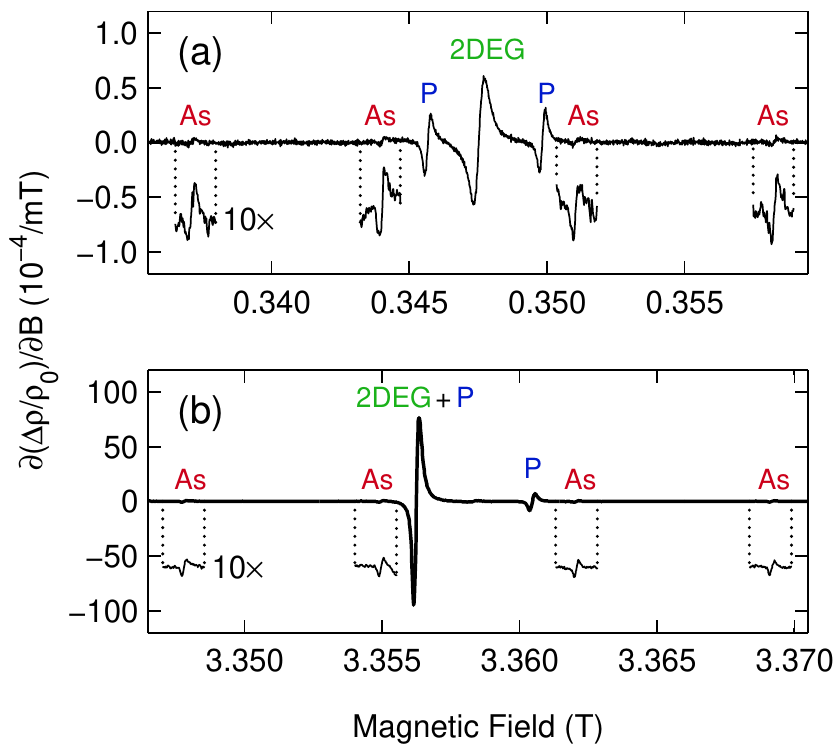}
\caption{\label{fig2}{(a) EDMR spectra obtained in X- and (b) W-band. The 2DEG, phosphorus (P) and arsenic (As) resonances are indicated along the traces. Sections of the EDMR spectra are magnified by 10$\times$ and offset for clarity. The gate bias was $0.3\:$V and the drain bias was $40\:$mV in both measurements.}}
\end{figure}

We carried out EDMR measurements in commercial Bruker ElexSys E680 X-band ($9.7\:$GHz) and W-band ($94\:$GHz) microwave resonators with corresponding Zeeman fields of $B=0.35\:$T and $3.36\:$T, respectively. A lock-in technique at $5.02\:$kHz and $0.2\:$mT field modulation was used to improve the signal-to-noise ratio. All measurements were carried out at $T=5\:$K where the device has a threshold voltage of $0.25\:$V and an effective mobility of $12\:000\:$cm$^2/$Vs. The Zeeman field is aligned in the plane of the 2DEG, perpendicular to the direction of current flow. No observable change in the sample current-voltage characteristics was seen under the different applied Zeeman fields. Further details of our measurement setup in W-band resonant microwave cavities including the design of our sample probe are discussed elsewhere \cite{lang10}. 

In all EDMR data discussed in this work, the microwave power was set to below the onset of microwave power broadening, where the signal reaches its maximum. 
Due to the use of magnetic field modulation, the spectra obtained are proportional to the first derivative of the change in device resistivity $\partial(\Delta \rho/\rho_0)/\partial B$, and typical spectra are shown in Fig.\:\ref{fig2}. 
We checked the signs of the signals and confirm that $\Delta\rho < 0$ on resonance in both X- and W-bands. Three groups of lines can be identified in the X-band spectrum (Fig.\:\ref{fig2}(a)). The intense center line has a $g$-factor of $g_{2DEG}=1.9999$ and is assigned to the 2DEG \cite{shankar07,shankar10}. The two adjacent peaks, split by a hyperfine coupling of $4.2\:$mT and with a center-of-gravity $g$-factor of $g_{P}=1.9987$, correspond to phosphorus donors with a nuclear spin of $1/2$ \cite{feher59}. Four smaller satellite peaks further out on both sides split by a hyperfine coupling of $7.1\:$mT arise from arsenic donors with a nuclear spin of $3/2$. The same three groups of lines are seen in the W-band spectrum (Fig.\:\ref{fig2}(b)), centered now at the high field of $3.358\:$T.  The field position of the 2DEG relative to the phosphorus center-of-gravity amounts to $\Delta B=h f_{\mu w}\//\mu_B(1/g_{2DEG}-1/g_{P})\approx-2.1\:$mT ($f_{\mu w}=94\:$GHz), and hence the 2DEG coincides with the low-field phosphorus line. This results in the two large resonance lines with different amplitudes around the center, while the four smaller hyperfine-split arsenic lines have equal amplitude. The resonance signals have peak-to-peak linewidths of $\sim 0.2\:$mT, and for a proper comparison of the inhomogeneously broadened resonance signals, we define the signal intensity of a resonance line as the area under the integrated spectrum. The signal intensities of the donor lines increase by a factor of $\sim20$ and the intensity of the 2DEG line by a factor of $\sim100$ from X-band to W-band. The ratios of the signal intensities are approximately 2DEG:P:As $=20:10:1$ in X-band and $100:10:1$ in W-band. The relative ratio between the phosphorus and arsenic signal intensities is consistent with the total number of dopants under the channel and also the number of hyperfine-split resonance lines.

\begin{figure}
\includegraphics[width=3.0 in]{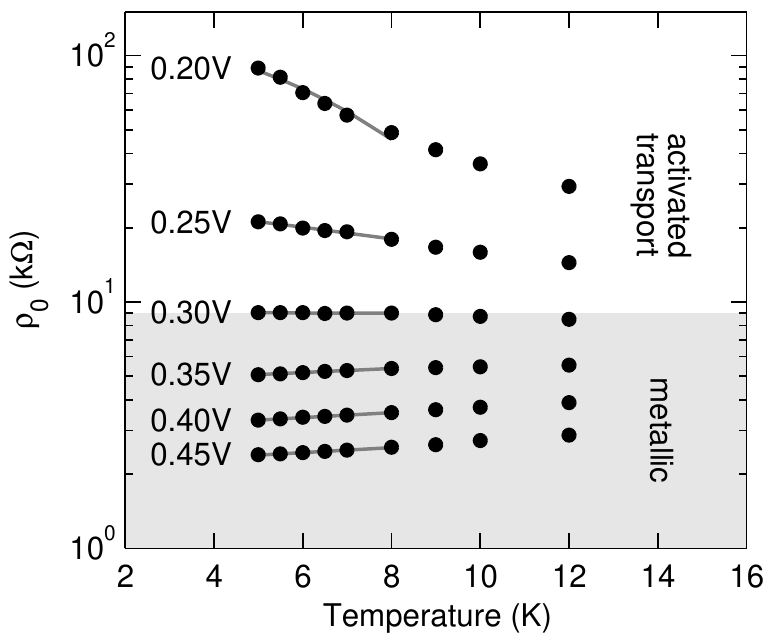}
\caption{\label{fig3}{Temperature dependence of device resistivity for $T=5-12\:$K on a semilogarithmic scale for gate voltages $V_g=0.25-0.45$V, as labeled. The shaded region designates the metallic transport regime. The lines corresponds to the linear best-fits of the temperature dependence of the sample resistivity for $T\leq8\:$K.}}
\end{figure}

In order to assess the possible contribution of bolometric heating of the 2DEG to the EDMR signal, we have measured the device resistivity over the temperature range $T=5-12\:$K as shown in Fig. \ref{fig3}. In this temperature range, acoustic phonon scattering does not contribute to the overall carrier mobility significantly \cite{kawaguchi82,ando82}, hence any temperature dependence of resistivity is a result of changes in $T_e$ only, independent of the lattice temperature $T_l$. We observe that carrier transport can be separated into two regimes: (i) $\partial\rho_0/\partial T < 0$ for $V_g < 0.3\:$V, the activated transport regime, and (ii) $\partial \rho_0/\partial T > 0$ for $V_g > 0.3\:$V, the metallic regime. For bolometric heating of the 2DEG one would expect the sign of the EDMR signal to follow the sign of the temperature gradient, therefore, the sign of the EDMR signal should change from negative to positive at around $V_g = 0.3\:$V. Our experiments show no such change in the sign of EDMR signal, and the sign disagrees with the temperature gradient ($\partial\rho_0 (T)/\partial T > 0$) for $V_g\geq0.3\:$V. Thus we conclude that bolometric heating does not produce any significant contribution to the EDMR signal.

Previous EDMR measurements of similar donor-doped MOSFETs at X-band have been attributed to spin-dependent neutral impurity scattering \cite{ghosh92, lo07}. More recently De Sousa {\it et al.} calculated the spin-dependent scattering cross sections, $\Sigma_s$ and $\Sigma_t$, within the framework of multi-valley effective mass theory \cite{desousa09}. They concluded that $\Sigma_s > \Sigma_t$, implying a positive EDMR signal, i.e. $\Delta\rho/\rho_0 > 0$. However, both Ghosh and Silsbee as well as our measurements show that $\Delta\rho/\rho_0 < 0$. While the experiments disagree with the sign predicted by the theory, we note that a refinement of the existing theory taking the full anisotropy of the silicon band structure into account might lead to cases where $\Sigma_s < \Sigma_t$ \cite{kwong91}. Another shortcoming of the spin-dependent scattering model is found in the ratio of the signal intensities of the 2DEG to the donors. 
The model predicts the 2DEG signal intensity to be equal to the sum of the hyperfine-split donor signal intensities, while our results show that the 2DEG signal intensity is much greater than the sum in both low- and high-field measurements. 
This can only be the case if spin-dependent scattering from paramagnetic centers other than neutral donors also contributes to the 2DEG signal. However, we have not observed resonance signals associated with surface defects such as dangling bonds (e.g. $P_b$ centers \cite{nishi71}). Finally, from the increase in thermal equilibrium polarizations we expect the spin-dependent scattering signal to be enhanced by a factor of $70$ at $T=5\:$K at W-band compared to X-band. The results in Fig.\:\ref{fig2} show that the 2DEG enhancement is stronger than expected, while the donor signal enhancement is substantially smaller. We have also measured similar EDMR spectra for different gate biases and the enhancement factors were similar. 
Due to these inconsistencies it is difficult to explain our results by invoking spin-dependent scattering alone. 

We therefore propose another EDMR mechanism, which originates from the polarization-dependent resistivity of the 2DEG \cite{graeff99, matsunami06} and a polarization transfer from the donor to the 2DEG electrons. We assume the 2DEG resistivity to be approximated by $\rho=\rho_1+\rho_2 p_c^2$, where $\rho_1$ is the polarization-independent background resistivity and $\rho_2$ the 2DEG polarization-dependent resistivity. 
Recognizing that $\rho_1 \gg \rho_2$ and assuming a complete saturation of the 2DEG spin transition, the 2DEG signal intensity is then predicted to be $\Delta\rho/\rho_0 \approx -p_{c}^2/(\rho_2/\rho_1)$. From the positive in-plane magnetoresistances ($\partial\rho/\partial B>0$), and hence positive correlations between 2DEG resistivity and spin polarization observed by others \cite{abrahams01,pudalov02,okamoto04}, we expect $\rho_2 > 0$. Thus, this model agrees with the negative sign of the EDMR signal observed in our experiments. At X-band, we expect $p_{c}\approx 1\%$ with the 2DEG densities used, and since $\Delta\rho/\rho_0\approx-10^{-5}$, we  conclude that $\rho_1/\rho_2\approx10$. Since $p_{c} \propto B$ for the 2DEG, the 2DEG signal should increase by 100 times from X- to W-band, which is indeed observed in our experiments. The EDMR signal intensities of the donor resonances depend on the effectiveness of the donor-to-2DEG polarization transfer, which is dominated by two quantities: (i) the spin relaxation rate of the 2DEG $1/T_{1c}$, and (ii) the spin exchange scattering rate $1/T_x$ \cite{Txfootnote}, which varies from donor to donor depending on their distance to the oxide interface as the 2DEG-donor wavefunction overlap changes \cite{desousa09} (we assume the spin relaxation rate of donors $1/T_{1d}$ to be much smaller than that of the conduction electrons as supported by electron paramagnetic resonance measurements \cite{shankar07,shankar10,schenkel06}). 
We first consider the limit $1/T_x\gg1/T_{1c}$, where the 2DEG and donor polarizations are strongly coupled and indistinguishable. In this case one would expect the 2DEG signal intensity to be equal to the donor signal intensities, which was not observed in our experiments. In the opposite limit where $1/T_x\ll1/T_{1c}$, the conduction electrons return to their thermal equilibrium rapidly, and hence the change in donor polarization on resonance has little effect on the 2DEG polarization. Therefore, no donor resonance signal is observed in this limit. 
The donor signals are most sensitive to donors with $1/T_x\sim1/T_{1c}$, and 
as $1/T_x$ is not expected to change much with magnetic field in the temperature range of our experiments \cite{desousa09b},the different 2DEG and donor signal intensity ratios at X- and W-band can therefore be explained if $1/T_{1c}$ becomes larger at higher magnetic fields: 
Donors with $1/T_x\sim1/T_{1c}$ at X-band will be less effective in influencing $p_c$ in W-band as now $1/T_x<1/T_{1c}$. This implies a reduced number of donors can contribute to the donor resonance signals in the high field measurements.
We are unaware of any experimental measurements of the magnetic field dependence of $1/T_{1c}$ in the metallic limit of a disordered 2DEG at this temperature range, and a direct measurement of the magnetic field dependence of $1/T_{1c}$ for the 2DEG will undoubtedly add new insights into the understanding of the origin of the EDMR effect in MOS devices. 

In conclusion, we have measured EDMR of n-type silicon field-effect transistors in X- and W-band microwave resonators, with corresponding Zeeman fields of $0.35\:$T and $3.36\:$T, respectively. Contrary to existing theoretical calculations for spin-dependent scattering between 2DEG and donor electrons, the sample resistivity was found to decrease on resonance. In addition, the 2DEG resonance signal showed much stronger magnetic field dependence than the donor resonance signals. These observations are consistent with a polarization-dependent 2DEG mobility model, where donors contribute to EDMR by a polarization transfer between the two spin systems. 

We thank Arzhang Ardavan, Rogerio de Sousa and Thorsten Last for useful discussions, and the UC Berkeley Microlab staff for technical support in device fabrication. 
This work was supported by the US National Security Agency under 100000080295. Additional supports by the Department of Energy under contract no DE-AC02-05CH11231 (LBNL), EPSRC through CAESR EP/D048559/1 (Oxford), and the National Science Foundation through the Princeton MRSEC under Grant No. DMR-0213706 (Princeton) are also acknowledged. V. L. is supported by Konrad-Adenauer-Stiftung e.V. and EPSRC DTA. J.J.L.M. is supported by The Royal Society and St. John's College, Oxford.



\begin{thebibliography}{25}
\bibitem{Txfootnote}
The spin exchange scattering rate per conduction electron should be $(n_d/n_c)/T_x$, where $n_d$ and $n_c$ are the donor and conduction electron densities, respectively. We abbreviate $(n_d/n_c)/T_x$ as $1/T_x$ in the main text for simplicity.
\bibitem{desousa09b}
See equation (11) of Ref. 7, where the donor $T_1^{-1}$ corresponds to the donor spin-flip exchange rate $T_x^{-1}$ discussed in this Letter. In the experimental conditions used, the exchange rate is not expected to increase by more than 10$\%$ from $B=0.35\:$T to $B=3.36\:$T.
\end{thebibliography}
\end{document}